\documentclass[journal]{IEEEtran}

\usepackage{amsmath}
\usepackage{cases}
\usepackage{amsfonts}
\usepackage{amssymb}
\usepackage{boxedminipage}
\usepackage{graphicx}
\usepackage[usenames]{color}
\usepackage{array,color}
\usepackage{colortbl}
\usepackage{lineno}
\usepackage{slashbox,pict2e}
\usepackage{stfloats}
\usepackage{algorithm}
\usepackage{algpseudocode}
\usepackage{multicol}
\usepackage{makecell}
\usepackage{caption}
\usepackage{subcaption}

\begin{document}

 \baselineskip=12pt
\title{ Downlink Resource Allocation for the High-speed Train and Local Users in OFDMA Systems}
\author{Chuang~Zhang$^\ast$,~Pingyi~Fan$^\ast$,~Ke~Xiong$^\dagger$\\  

$^\ast$State Key Laboratory on Microwave and Digital Communications\\
Tsinghua National Laboratory for Information Science and Technology\\
Department of Electronic Engineering, Tsinghua University, Beijing 100084, China
\\
$^\dagger$School of Computer and Information Technology, Beijing Jiaotong University, Beijing, P.R. China\\
E-mail:~zhangchuang11@mails.tsinghua.edu.cn,~fpy@tsinghua.edu.cn,~kxiong@bjtu.edu.cn}
\maketitle

\begin{abstract}
We consider providing services for passengers in a high-speed train and local users (quasi-static users) in a single OFDMA system. For the train, we apply a two-hop architecture, under which, passengers communicate with base stations (BSs) via a mobile relay (MR) installed in the train cabin. With this architecture, all passengers in the train can be represented by the MR. Since the channels of the MR and local users vary differently, we consider allocating system resources (power and subcarriers) over two time-scales for them. We formulate the problem as a capacity optimization problem for the MR subject to the sum capacity constraint of local users. We treat the inter-carrier interference (ICI) at the MR as additive Gaussian noise and derive an explicit expression for the ICI using the two-path Doppler spread model. Then we discuss the optimization problem and propose an optimal power and subcarrier allocation (OPSA) policy. The capacity obtained using OPSA is compared with that of constant power and subcarrier allocation (CPSA) policies. Simulation results justify the optimality of the OPSA. Besides, by comparing the capacity bounds achieved by OPSA with and without ICI, we find that only in specific regions, where the gap between the capacity bounds is large, do practical ICI cancellation methods provide meaningful rate gain.
\end{abstract}


\section{Introduction}

Traditional resource allocations of OFDMA systems unanimously assume that all users in the system are quasi-static (or in low-mobility). However, with the rapid development of group transportation systems like high-speed railways in recent years, there are scenarios in which groups of high-mobility users pass through an area and desire service from the same BS. To provide broadband wireless services for both high-mobility users and local quasi-static users (local users for short), it is necessary to reconsider the resource allocation problem since the channel conditions of high-mobility users and local users are quite different.

We investigate downlink resource allocation of an OFDMA system with a high-speed train and many local users. We employ the two-hop architecture as in \cite{wang2012dasmchst} \cite{zhang2013sbhsrbsa} for the train and treat all train passengers as one big user denoted by the MR. This scenario has the following distinctions with traditional ones. First, since the train usually transports several hundreds of people, the MR requires a much larger amount of resources than individual local user. Second, the instantaneous channel state information (CSI) of the MR is not available to the BS due to rapid time variations. Third, the MR is subject to ICI due to severe Doppler spread, while the ICI of local users can be ignored. Considering these differences, we allocate power and subcarriers over two time-scales for them, and we aim at maximizing the capacity of the MR while satisfying the sum capacity constraint of local users.

Our main contributions are as follows: \emph{First}, we formulate the joint power and subcarrier allocation problem for downlink OFDMA system in the scenario with a passing high-speed train and many local users. \emph{Second}, using the two-path Doppler spread model, we derive a closed-form expression for the ICI at the MR. Based on the expression, we transform the resource allocation problem into a convex optimization problem, and prove the existence and uniqueness of optimal solutions. Furthermore, we propose an efficient algorithm to obtain the optimal solution. \emph{Third}, by comparing the achievable capacities of OPSA with and withou ICI, we provide useful guidance for the application of practical ICI cancellation methods.

The rest of the paper is organized as follows. In Section \ref{sec_systmod}, the system model is introduced. In Section \ref{sec_prosetup} and \ref{sect_protrans}, we setup the optimization problem and make a transformation to it, respectively. In Section \ref{sec_sop}, the optimization problem is solved based on convexity analysis. In Section \ref{sec_simulation}, simulation results are provided to justify the analysis. Finally, conclusion is given in Section \ref{sec_conclusion}.

\section{System Model}\label{sec_systmod}
\begin{figure}
  \centering
  \includegraphics[width=0.45\textwidth]{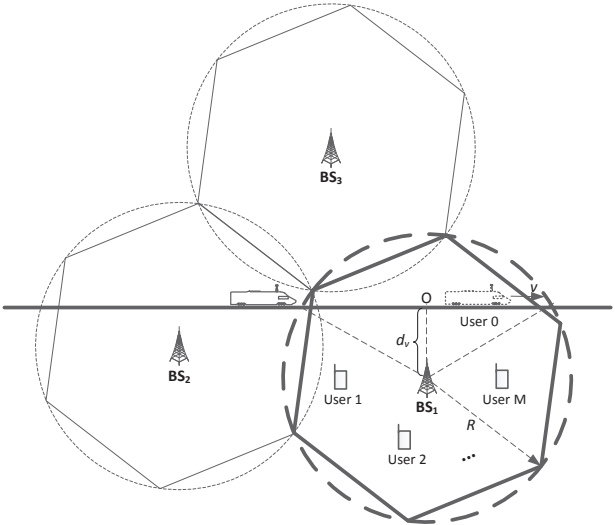}\\
  \caption{Network model.}\label{fig_sysmodel}
\end{figure}


\subsection{Network Model}
The network model is shown in Fig. \ref{fig_sysmodel}. In this system, there are $M$ local users (moving velocity less than 10 km/h and denoted by user $m=1,2,\ldots,M$) and a passing high-speed train (moving velocity larger than 200 km/h, and denoted by user $0$). The radius of the coverage area of a BS is $R$ m. The bandwidth of the system is $B$ Hz, which is divided into $N$ subcarriers, and denoted as $\Omega=\{0,1,\ldots, N-1\}$. The available power of the BS is $P$ W. We study resource allocation in one cell, and ignore interference caused by neighboring cells. Since the boundary areas between two cells are usually small, for the convenience of analysis, we assume that one BS serves a circular area, as shown in Fig. \ref{fig_sysmodel}.

The railway is located at a vertical distance $d_v$ from $\text{BS}_1$. The train travels with a constant velocity $v$ when passing through the cellular area of $\text{BS}_1$. Set the axis along the railway and let $O$ in Fig. \ref{fig_sysmodel} be the origin. Let the time when the train passes point $O$ be time zero. Then from $-\frac{T_s}{2}$ to $\frac{T_s}{2}$, where $T_s=\frac{d_s}{v}$, $d_s=2\sqrt{R^2-d_v^2}$, the train is in the coverage area of $\text{BS}_1$. We then consider allocating the resources of $\text{BS}_1$ to both the train and local users in this period of time.

\subsection{Channel Model} \label{sec_chmod}
\subsubsection{Large-Scale Fading}


Since local users in the system do not have much geographic movement in a short period, they can be considered as stationary in the period $[-\frac{T_s}{2},\frac{T_s}{2}]$. Thus, their average path losses can be treated as invariant. Suppose user $m$, $m=1,2,\ldots,M$, is located at a distance $d_m$ from the BS, then the path loss for user $m$ is $L_m(d_m)=10\alpha\log(d_m)$, where $\alpha$ is the path loss exponent.

A time-varying path loss is experienced by the MR in the train. Let the distance between the MR and $\text{BS}_1$ at time $t$ be $d_0(t)=\sqrt{d_v^2+(vt)^2}$. Then, the corresponding path loss is $L_0(d_0(t))=5\alpha\log(d_v^2+(vt)^2)$.

\subsubsection{Small-scale Fading}


For user $m$, $m=0,1,\cdots,M$, the CSI at time $t$ is
\begin{equation}
h_m(t,f)=\sum\nolimits_{\tilde{l}=0}^{\tilde{L}-1}h_{m\tilde{l}}(t)e^{-j2\pi f\tau_{m\tilde{l}}},
\end{equation}
where $h_{m\tilde{l}}(t)$ and $\tau_{m\tilde{l}}$ are the complex amplitude and delay of the $\tilde{l}$th path of user $m$ respectively, and $\tau_{m0}\leq\tau_{m1}\leq\cdots\leq\tau_{m\tilde{L}-1}$, $f$ is the frequency. The multipath fading for all users is with the same exponential delay profile, i.e., $S(\tau)=\frac{1}{\sigma} e^{-\tau/\sigma}$, where $\sigma$ is the average delay.

For local users, their Doppler spread is zero. For the MR, we apply the two-path Doppler spread model \cite{robertson1999alotd} \cite{li2001botiiofdmtvi} for each tap. This model is utilized here for the following reasons. First, with the simplified two-path model, we can obtain a closed-form expression for the ICI term, and this helps us to analyze the effect of Doppler spread on the resource allocation problem more clearly. Second, the time-domain autocorrelation function of the two-path model is a cosine function, and the autocorrelation function of other Doppler spread model, like Jakes model or Rice model, can be expressed in the form of cosine series. With finite term approximation, we can obtain similar results for these general models as well.

Let $f_D$ be the Doppler shift, then the Doppler spectrum of the two-path model is
\begin{align}\label{eqn_Dosp2path}
P(f)=\tfrac{1}{2}(\delta(f+f_D)+\delta(f-f_D)).
\end{align}
The corresponding time-domain autocorrelation is
\begin{align}\label{eqn_2pathautocorr}
R(t)=\cos(2\pi f_D t).
\end{align}

Similar to \cite{wang2006pdofdmos} \cite{hamdi2011ueraofdmtvc}, we assume that the CSI of all users are wide sense stationary and $h_m(t,f)$ are zero-mean complex Gaussian processes. Then, based on the above assumptions, for local users, $h_m(t,f)$ has cross covariance function
\begin{align}
R_m(t_1,f_1;t_2,f_2)&=\tfrac{1+j2\pi\sigma(f_2-f_1)}{1+[2\pi\sigma(f_2-f_1)]^2},\; m=1,2,\ldots,M.\nonumber
\end{align}
For the MR, $h_0(t,f)$ has cross covariance function
\begin{align}
R_0(t_1,f_1;t_2,f_2)=\tfrac{1+j2\pi\sigma(f_2-f_1)}{1+[2\pi\sigma(f_2-f_1)]^2}\cos(2\pi f_D(t_2-t_1)).\nonumber
\end{align}


\subsection{Transmission  Model}

The baseband OFDM block at the BS can be expressed as
\begin{align}
x(t)=\frac{1}{\sqrt{N}}\sum\nolimits_{n=0}^{N-1}s_ne^{j2\pi\frac{n}{T}t}, \ t\in[-T_{\text{CP}},T],
\end{align}
where $N$ is the number of subcarriers, $s_n$ is the symbol on subcarrier $n$, $T_{\text{CP}}$ is the length of cyclic prefix, and $T$ is the length of effective data symbol.

We assume that perfect synchronization can be obtained at each user. Then, for user $m$, $m=0,1,\cdots,M$, the received signal after removing the cyclic prefix is
\begin{align} \nonumber
y_m(t)=\frac{1}{\sqrt{N}}\sum\nolimits_{n=0}^{N-1}s_n h_{m}\Big(t,\frac{n}{T}\Big)e^{j2\pi\frac{n}{T}t}+w(t), \, t\in[0, T],
\end{align}
where $w(t)$ is the additive noise.

After A/D conversion and FFT, the received signal on the $p$th subcarrier for user $m$ is given by
\begin{align} \label{eqn_recfft}
y_m(p)=&\frac{1}{\sqrt{N}}\sum\nolimits_{k=0}^{N-1}y_m\Big(k\frac{T}{N}\Big)e^{-j\frac{2\pi}{N}kp},  \nonumber\\
=&\frac{1}{N}\sum\nolimits_{k=0}^{N-1}\sum\nolimits_{n=0}^{N-1}s_nh_m\Big(k\frac{T}{N},\frac{n}{T}\Big)e^{j\frac{2\pi}{N}(n-p)k}\nonumber\\
&+w(p), \quad p=0,1,\ldots, N-1
\end{align}
where $w(p)=\frac{1}{\sqrt{N}}\sum_{k=0}^{N-1}w(k\frac{T}{N})e^{-j\frac{2\pi}{N}kp}$, and is assumed to be complex Gaussian noise with density $N_0$.

By defining
\begin{align}\label{eqn_ICI}
H_m(n,p)=\frac{1}{N}\sum\nolimits_{k=0}^{N-1}h_m\Big(k\frac{T}{N},\frac{n}{T}\Big)e^{j\frac{2\pi}{N}(n-p)k},
\end{align}
we can further express   (\ref{eqn_recfft}) as
\begin{align}\label{eqn_Umreceiv}
y_m(p)&=s_pH_m(p,p)+\sum_{n=0,n\neq p}^{N-1}s_n H_m(n,p)+w(p).
\end{align}

\section{Problem Setup} \label{sec_prosetup}

In this section, we set up our optimization problem based on two time-scale resource allocation.

\subsection{Resource Allocation over Two Time-scales}

The instantaneous CSI of the MR is not available to the BS. Therefore, resource allocation for the MR can only be done based on the average CSI, which is equivalent to utilizing large-scale fading. For local users, the instantaneous CSI can be accurately tracked and resource allocation for them can be done based on CSI of small-scale fading. Considering such differences, we let the system operate in two time-scales, a longer time-scale on the order of seconds and a shorter time-scale on the order of miliseconds. In the longer time-scale, the BS divides the power and subcarriers between the MR and local users. In the shorter time-scale, the power and subcarriers allocated to local users are further allocated to them according to their instantaneous CSI. We use scheduling period to denote the longer time-scale and slot to denote the shorter time scale.

Assume each scheduling period lasts $\tau_l$ seconds, and each slot lasts $\tau_s$ seconds, $\tau_s<<\tau_l$. Each scheduling period has $L$ slots, i.e., $L=\lceil\frac{\tau_l}{\tau_s}\rceil$. The time period $[-\frac{T_s}{2},\frac{T_s}{2}]$ is divided into $2I+1$ scheduling periods, where  $I=\lceil\frac{T_s}{2\tau_l}\rceil$.

At the beginning of each scheduling period, the BS divides its power $P$ and subcarriers $\Omega$ into two parts. Let $\Omega_{0}(i)$, $P_{0}(i)$ be the set of subcarriers and power allocated to the MR in the $i$th scheduling period, respectively. $\Omega_U(i)=\Omega/\Omega_0(i)$, $P_U(i)=P-P_0(i)$ be the set of subcarriers and power allocated to local users, respectively.  This separation remains the same in the whole scheduling period. In each slot, $\{P_U(i),\Omega_U(i)\}$ is further assigned among local users based on their instantaneous CSI.

In the $l$th slot of each scheduling period, $l=1, 2,\ldots,L$, the BS allocates power $P_U$ and subcarriers $\Omega_U$ among local users according to $\{P_{m,p}(l),b_{m,p}(l)\}$, where $b_{m,p}(l)$ is a binary indicator that $b_{m,p}(l)=1$ if subcarrier $p$ is allocated to user $m$, and $b_{m,p}(l)=0$ otherwise. $P_{m,p}(l)$ is the power allocated to user $m$ on subcarrier $p$. Then $\{P_{m,p}(l), b_{m,p}(l)\}$ have $\sum_{m=1}^{M}b_{m,p}(l)\leq 1, \; p\in\Omega_U$ (one subcarrier is only allocated to one user), $\sum_{m=1}^{M}\sum_{p \in \Omega_U}P_{m,p}(l)\leq P_U$ (the sum power of all subcarriers should be no greater than the power allocated to local users).

\subsection{Capacity of Local Users and the MR}

It can be verified that when the Doppler spread $f_D=0$, the channel $h_m(t,f)$ becomes time-invariant in an OFDM block, and $H_m(n,p)=0$ for $n\neq p$. Therefore, for local users, \begin{align}
y_m(p)&=s_pH_m(p,p)+w(p),\ m=1,2,\ldots,M. \nonumber
\end{align}

Furthermore, if we use one-tap equalizer and assume perfect estimation of CSI at the receiver, the capacity of user $m$ in the $l$th slot of the $i$th scheduling period is
\small
\begin{align}
& C_{m}(i_l)= \sum_{p\in\Omega_U(i)}\frac{B b_{m,p}(l)}{N}\log\bigg( 1+\frac{|H_m(p,p)(l)|^2P_{m,p}(l)}{d_m^\alpha N_0B/N}\bigg),\nonumber
\end{align}
\normalsize
where $m=1,2,\cdots,M$ and $\Omega_U(i)$, $b_{m,p}(l)$ and $P_{m,p}(l)$ have the same notations as before. $H_m(p,p)(l)$ is $H_m(p,p)$ in the $l$th slot, $d_m^\alpha$ is the path loss. Besides, implicit in the capacity definition is a block fading assumption that each slot can be treated as a block.

Different from local users, the MR is subject to severe Doppler spread. Therefore, $H_m(n,p)\neq 0, n\neq p$. By treating the interference in   (\ref{eqn_Umreceiv}) as Gaussian noise, the capacity of the MR in the $l$th slot of the $i$th scheduling period is
\begin{align} 
C_{0}(i_l)=\sum_{p\in\Omega_0(i)}\frac{B}{N}E\bigg(\log\bigg( 1+\tfrac{\frac{|H_0(p,p)(l)|^2P_{0,p}(l)}{(d_v^2+(vi\tau_l)^2)^{\frac{\alpha}{2}}}}{P_{\text{ICI}}(l)+N_0B/N}\bigg)\bigg),\nonumber
\end{align}
where
\begin{align}\label{eqn_ICIterm}
P_{\text{ICI}}(l)=&\sum\nolimits_{n\in\Omega_0(i),n\neq p}\tfrac{P_{0,n}(l) |H_0(n,p)(l)|^2}{(d_v^2+(vi\tau_l)^2)^{\frac{\alpha}{2}}} + \\
& \sum\nolimits_{m=1}^{M}\sum\nolimits_{n\in\Omega_U(i)} \tfrac{b_{m,n}(l)P_{m,n}(l) |H_0(n,p)(l)|^2}{(d_v^2+(vi\tau_l)^2)^{\frac{\alpha}{2}}} \nonumber
\end{align}
is the ICI and expectation is over $|H_0(p,p)(l)|^2$. Since the channel varies fast in each slot, the capacity of the MR is approximated by the ergodic capacity. Note that the path loss in a scheduling period is invariant.

\subsection{Problem Formulation}

In each scheduling period, we allocate power and subcarriers to maximize the capacity of the MR while guaranteeing that the sum capacity of local users in this period is no less than a threshold. Then, this problem is formulated as \textbf{P1:}
\begin{align}
\max_{\{P_0(i),\Omega_0(i)\}} & \; \tfrac{1}{L}\sum\nolimits_{l=1}^{L} C_0(i_l) \label{eqn_mainopt} \\
\text{s.t.}   & \; \tfrac{1}{L}\sum\nolimits_{l=1}^{L}\sum\nolimits_{m=1}^{M}C_{m}(i_l)\geq R_{\text{th}}, \tag{\theequation a}\label{eqn_mainopta} \\
& \; \Omega_U(i)\cup\Omega_{0}(i)\subseteq \Omega,\; \ P_U(i)+P_{0}(i)\leq P.\tag{\theequation b}\label{eqn_mainoptb}
\end{align}

\section{Problem Transformation} \label{sect_protrans}

In this section, we discuss resource allocation schemes in a slot and make an equivalent transformation to problem \textbf{P1}.
\subsection{Resource Allocation for Local Users}

It was proven in \cite{jang2003tpamofdms} that the optimal power and subcarrier allocation policy to maximize the sum capacity of an OFDMA system is first to select for each subcarrier the user with the best CSI and then allocate the power using water-filling among the subcarriers. However, this approach can be quite unfair when users are located at different positions. So in order to maintain a certain degree of fairness and at the same time to utilize the variations of CSI, we propose a scheme that after detecting the CSI of each local user, the BS multiply the CSI by the corresponding path loss of that user to mitigate the effect of large-scale fading. Then the BS allocates power and subcarriers solely based on small scale fading. It can be conceived that in a longer time period (e.g., a scheduling period), all users would approximately have a fair share of resources since they have the same small-scale fading statistics. Besides, it was noted in \cite{jang2003tpamofdms} that when the number of users in the system is large, simple equal power allocation among subcarriers can obtain performance with marginal difference with water-filling. Since a large number of users is assumed in the system, we adopt the simple equal power allocation to reduce computational complexity.

Then power and subcarrier allocation for local users is first to choose the user with the best small-scale fading CSI for each subcarrier $p$, i.e., allocating subcarrier $p$ to user $\arg \max_{m} |H_m(p,p)(l)|$ (choose the user with the smallest index when more than one user has the same largest CSI), and then to allocate the power equally among the subcarriers.

In different slots, the statistics of CSI are the same, we omit the slot index in $H_m(p,p)(l)$. Let $|H_{\text{max}}(p)|$ denote the random variable $|H_{\text{max}}(p)|=\max_{m}\,|H_m(p,p)|$. Furthermore, the channel statistics of each subcarrier is the same in our model. Then when the number of slots $L$ is large in a scheduling period, the sum capacity of local users with all system resources can be approximated by the statistical average
\begin{align}
C_{\text{sum}}=\sum_{m=1}^{M}\frac{N}{M}\frac{B}{N}E\bigg(\log\bigg( 1+\frac{|H_{\text{max}}(p)|^2P/N}{d_m^\alpha N_0B/N}\bigg)\bigg)\nonumber
\end{align}

We set the threshold $R_{\text{th}}$ in Inequlity (\ref{eqn_mainopta}) as a ratio of $C_{\text{sum}}$, that is, $\rho C_{\text{sum}}$, where $0\leq \rho\leq 1$. Similarly, the left side of constraint (\ref{eqn_mainopta}) becomes
\begin{align}\nonumber
\sum_{m=1}^{M}\frac{|\Omega_U(i)|}{M}\frac{B}{N}E\bigg(\log\bigg( 1+\tfrac{\frac{|H_{\text{max}}(p)|^2P_U(i)}{d_m^\alpha|\Omega_U(i)|}}{ N_0B/N}\bigg)\bigg).
\end{align}

\subsection{Resource Allocation for the MR}

Since the BS can not obtain the instantaneous CSI of the MR, it allocates the power equally among the subcarriers for the MR. The fading statistics of each subcarrier in different slots at the MR are the same, so we can omit the slot index in the channel gain $H_0(p,p)(l)$ as well. Furthermore, the fading statistics of different subcarriers are also the same. Then, in each scheduling period, the time average of (\ref{eqn_mainopt}) can be approximated by the statistical average
\begin{align} \nonumber 
C_{0}(i)=\frac{B|\Omega_0(i)|}{N}E\bigg(\log\bigg( 1+\tfrac{\frac{|H_0(p,p)|^2P_0(i)}{(d_v^2+(vi\tau_l)^2)^{\frac{\alpha}{2}}|\Omega_0(i)|}}{P_{\text{ICI}}(i)+N_0B/N}\bigg)\bigg),
\end{align}
the ICI term then becomes

\small
\begin{flalign}\label{eqn_ICIterm_sim}
P_{\text{ICI}}(i)=\sum\limits_{\scriptstyle n\in\Omega_0(i),\hfill\atop
\scriptstyle n\neq p \hfill} \tfrac{\frac{P_0(i)|H_0(n,p)|^2}{|\Omega_0(i)|}}{(d_v^2+(vi\tau_l)^2)^{\frac{\alpha}{2}}}+ \sum_{n\in\Omega_U(i)} \tfrac{\frac{P_U(i) |H_0(n,p)|^2}{|\Omega_U(i)|}}{(d_v^2+(vi\tau_l)^2)^{\frac{\alpha}{2}}}.
\end{flalign}
\normalsize

Then, problem (\ref{eqn_mainopt}) can be equivalently expressed as \textbf{P2:}
\small
\begin{align}
\max_{\{P_{0}(i),\Omega_{0}(i)\}}  & \sum_{p\in \Omega_{0}(i)}\frac{B }{N}E\bigg(\log\bigg( 1+\tfrac{\frac{|H_0(p,p)|^2P_{0}(i)}{(d_v^2+(vi\tau_l)^2)^{\frac{\alpha}{2}}|\Omega_{0}(i)|}}{P_{\text{ICI}}(i)+\frac{N_0B}{N}}\bigg)\bigg) \label{eqn_simmainopt}\\
\text{s.t.}\quad  & \sum_{m=1}^{M}\frac{|\Omega_U(i)|}{M}\frac{B}{N}E\bigg(\log\bigg( 1+\tfrac{\frac{|H_{\text{max}}(p)|^2P_U(i)}{d_m^\alpha|\Omega_U(i)|}}{ N_0\frac{B}{N}}\bigg)\bigg)\geq R_{\text{th}}, \tag{\theequation a}\label{eqn_simmainopta}\\
  & \Omega_U(i)\cup\Omega_{0}(i)\subseteq \Omega,\ P_U(i)+P_{0}(i)\leq P, \tag{\theequation b}\label{eqn_simmainoptb}
\end{align}
\normalsize

\section{Solving the Optimization Problem} \label{sec_sop}

In this section, we solve the optimization problem (\ref{eqn_simmainopt}). To do so, we first derive an explicit expression for the ICI term.


The ICI of subcarrier $n$ on subcarrier $p$ is $E(H(n,p)H^*(n,p))$. Substituting (\ref{eqn_ICI}) into the expression and after complex derivations, we get the form of (\ref{eqn_ICI2path}).
\small
\begin{figure*}[hb]
\hrulefill
\begin{align}\label{eqn_ICI2path}
E(H(n,p)H^*(n,p))=\tfrac{1}{N}+\tfrac{2}{N^2}\sum_{k=1}^{N-1}\tfrac{N-k}{2}\Big(\cos\Big((2\pi f_D(i)\tfrac{T}{N} +\tfrac{2\pi}{N}(n-p))k\Big)+\cos\Big((2\pi f_D(i)\tfrac{T}{N} -\tfrac{2\pi}{N}(n-p))k\Big)\Big).
\end{align}
\end{figure*}
\normalsize

Note $\mathbb{A}=2\pi f_D(i)\frac{T}{N} +\frac{2\pi}{N}(n-p)$, $\mathbb{B}=2\pi f_D(i)\frac{T}{N} -\frac{2\pi}{N}(n-p)$, then (\ref{eqn_ICI2path}) can be expressed as
\small
\begin{align}\label{eqn_ICIcal}
E(H(n,p)H^*(n,p))=\frac{1}{2N^2}\bigg(\frac{\sin^2(\frac{\mathbb{A}N}{2} )}{\sin^2(\frac{\mathbb{A}}{2})}+\frac{\sin^2(\frac{\mathbb{B}N}{2})}{\sin^2(\frac{\mathbb{B}}{2})}  \bigg),
\end{align}
\normalsize%

In our discussions, $f_D(i)T\ll 1$, therefore, $\frac{\pi}{N}f_D(i)T\ll 1$, and $ \frac{\pi}{N}f_D(i)T\ll \frac{\pi}{N}|n-p|$. Thus, the term $\frac{\pi}{N}f_D(i)T$ in the denominator of  (\ref{eqn_ICIcal}) can be ommited. As a result,
\begin{align} \label{eqn_ICIclose}
E(H(n,p)H^*(n,p))=\frac{1}{N^2}\frac{\sin^2(\pi f_D(i)T)}{\sin^2(\frac{\pi}{N}(n-p))}.
\end{align}
(\ref{eqn_ICIclose}) is the explicit expression for the ICI term. We see from this equation that the ICI increases with the Doppler spread, the number of subcarriers, and decreases with the index difference between subcarriers. Considering the time-varying Doppler shift in the high-speed railway scenario,
\begin{align} 
f_D(t)=\frac{vf_c}{c}\frac{vt}{\sqrt{d_0^2+(vt)^2}},\ -\frac{d_s}{2v}\leq t\leq \frac{d_s}{2v}, \nonumber
\end{align}
we see that the ICI of the MR is also time-varying. 

Substituting   (\ref{eqn_ICIclose}) into   (\ref{eqn_ICIterm_sim}), we can get the ICI as
\small
\begin{align}  \label{eqn_ICIofallexp}
P_{\text{ICI}}(i)=&\tfrac{\sin^2(\pi f_D(i)T)P_0(i)}{(d_0^2+(vi\tau_l)^2)^{\frac{\alpha}{2}}N^2|\Omega_0(i)|}\sum_{n\in\Omega_0(i),n\neq q }\tfrac{1}{\sin^2(\frac{\pi}{N}(n-p))} \nonumber\\
&+\tfrac{\sin^2(\pi f_D(i)T)P_U(i)}{(d_0^2+(vi\tau_l)^2)^{\frac{\alpha}{2}}N^2|\Omega_U(i)|}\sum_{n\in\Omega_U(i) }\tfrac{1}{\sin^2(\frac{\pi}{N}(n-p))}.
\end{align}
\normalsize

We see from   (\ref{eqn_ICIclose}) that when the index difference between two subcarriers is large, the ICI would be quite small. Therefore, only the ICI of several neighboring subcarriers should be considered. From our simulation results, we find that when the subcarrier index difference is larger than $5$, the ICI is quite small and could be ignored. Besides, since the subcarriers allocated to the MR are consecutive, most of them would experience the same ICI induced by subcarriers allocated to the MR except for some boundary subcarriers. Then we could no longer consider the second term in   (\ref{eqn_ICIofallexp}). Thus, the objective function of problem (\ref{eqn_mainopt}) becomes
\begin{align}
\frac{B|\Omega_{0}(i)|}{N}E\bigg(\log\bigg( 1+\tfrac{\frac{|H_0(p,p)|^2P_{0}(i)}{(d_v^2+(vi\tau_l)^2)^{\frac{\alpha}{2}}|\Omega_{0}(i)|}}{P_{\text{ICI}_0}(i) +N_0B/N}\bigg)\bigg),
\end{align}
where $P_{\text{ICI}_0}(i)$ is the first term of   (\ref{eqn_ICIofallexp}).

Since the number of subcarriers in the system is large, we can use a continuous approximation for the problem. Let $P_{0}(i)=\eta(i)P$, $|\Omega_{0}(i)|=\beta(i)N$, $\gamma_0(i)=\frac{|H_0(p,p)|^2P}{(d_v^2+(vi\tau_l)^2)^{\frac{\alpha}{2}}N_0B}$, $\gamma_m(i)=\frac{|H_{\text{max}}(p)|^2P}{d_m^\alpha N_0B}$, $\gamma_{\text{ICI}_0}(i)=\frac{\sin^2(\pi f_D(i)T)P}{N^2(d_v^2+(vi\tau_l)^2)^{\frac{\alpha}{2}}N_0B} \sum_{n \in\Omega_{0}(i) }\frac{1}{\sin^2(\frac{\pi}{N}(n-p))}$, then problem (\ref{eqn_simmainopt}) becomes
\small
\begin{align}
\max_{\{\eta(i),\beta(i)\}} & \; \beta(i) B E\bigg(\log\bigg( 1+\frac{\gamma_0(i)\eta(i)}{\gamma_{\text{ICI}_0}(i)\eta(i)+\beta(i)}\bigg)\bigg)\label{eqn_finalopt} \\
\text{s.t.}& \;  \sum_{m=1}^{M}(1-\beta(i))\frac{B}{M} E\bigg(\log\bigg( 1+\gamma_m(i)\frac{1-\eta(i)}{1-\beta(i)}\bigg)\bigg)\geq R_{\text{th}}, \tag{\theequation a} \label{eqn_finalopta}\\
  &\; 0\leq \beta(i)\leq 1,\ 0\leq\eta(i)\leq 1, \tag{\theequation b} \label{eqn_finaloptb}
\end{align}
\normalsize
Regarding problem (\ref{eqn_finalopt}), we have

\textbf{Theorem 1:} The objective function of problem (\ref{eqn_finalopt}) is concave with respect to (w.r.t.) the vector $\{\eta(i),\beta(i)\}$.
\begin{proof}
We see from  \cite{boyd2004co} Section 3.2.6 that perspective operation preserves convexity. For a function $f: R^{n}\rightarrow R$, and its perspective function $g(x,t):R^{n+1}\rightarrow R$ defined with $ g(x,t)=tf(x/t)$, $\text{dom}\,g=\{(x,t)|x/t\in \text{dom}\, f \}.$ If $f(x)$ is convex w.r.t. $x$, then $g(x,t)$ is convex w.r.t. $\{x,t\}$. This also applies for concave functions.

It can be easily seen that the function $ f(\eta(i))= B E\Big(\log\Big( 1+\tfrac{\gamma_0(i)\eta(i)}{\gamma_{\text{ICI}_0}(i)\eta(i)+1}\Big)\Big)$ is a concave function of $\eta(i)$. Define $g(\eta(i),\beta(i))=\beta(i) B E\Big(\log\Big( 1+\tfrac{\gamma_0(i)\eta(i)}{\gamma_{\text{ICI}_0}(i)\eta(i)+\beta(i)}\Big)\Big)$, then $g(\eta(i),\beta(i))$ is derived from the perspective operation of $f(\eta(i))$ and $\frac{\eta(i)}{\beta(i)}\in \text{dom} f(\eta(i))$.

According to the above discussions,  $g(\eta(i),\beta(i))$ is a concave function w.r.t. the vector $\{\eta(i),\beta(i)\}$, i.e., the objective function of  problem (\ref{eqn_finalopt}) is concave.
\end{proof}

Based on Theorem 1, we can conclude that

\textbf{Corollary 1:} The optimization problem (\ref{eqn_finalopt}) has a unique optimal solution.



Although we prove the existence and uniqueness of optimal solutions of problem (\ref{eqn_finalopt}), there is no explicit expression for the solution. Here, we devise an algorithm to obtain the optimal solution based on the concave property of the objective function. Let
$G_1(\eta(i),\beta(i))=\beta(i) B E\Big(\log\Big( 1+\gamma_0(i)\frac{\eta(i)}{\beta(i)}\Big)\Big)$,
$G_2(\eta(i),\beta(i))=\sum_{m=1}^{M}(1-\beta(i))\frac{B}{M} E\bigg(\log\bigg( 1+\gamma_m(i)\frac{1-\eta(i)}{1-\beta(i)}\bigg)\bigg)$,
then the algorithm is presented as Alg. \ref{alg_optsol}.

\floatname{algorithm}{Alg.}
\begin{algorithm}
\caption{Optimal Power and Subcarrier Allocation (OPSA)} \label{alg_optsol}
\begin{algorithmic}[1]
\State \textbf{input:} $d_v, v, \alpha, i, P, B, N_0, \tau_l, R_{\text{th}}, \text{statistics of} \ H_0(p,p)(i) $, set initial values for $\eta(i)$, $\beta(i)$, set value for $\beta_{\text{sp}}$,
\State calculate $G_1(\eta(i),\beta(i))$, $G_2(\eta(i),\beta(i))$,
\State $C_{\text{trm}}\leftarrow G_1(\eta(i),\beta(i))$,
\While {$G_1(\eta(i),\beta(i))>=C_{\text{trm}}$ }
\State $C_{\text{trm}}\leftarrow G_1(\eta(i),\beta(i))$,
\State $\beta(i)\leftarrow \beta(i)+\beta_{\text{sp}}$,
\State calculate $ \eta(i)$ which makes $G_2(\eta(i),\beta(i))=R_{\text{th}}$ using methods like dichotomy,
\State $\eta(i)\leftarrow \eta(i)$,
\EndWhile
\State \textbf{output:} $C_{\text{trm}}$, $\beta(i)$, $\eta(i)$
\end{algorithmic}
\end{algorithm}



\section{Simulation Results}\label{sec_simulation}

In this section, we conduct simulations to confirm the analytical results. Common parameters are in TABLE \ref{tab_parset}. Tap delays are in TABLE \ref{tab_tdelay}. Besides, $50$ users are divided into five groups equally, and each group has the same distance from the BS, which are $100$m, $1325$m, $2550$m, $3775$m, and $5000$m, respectively.

%

\begin{table}[!ht]
\setlength{\abovecaptionskip}{2pt}
\setlength{\belowcaptionskip}{0pt}
\caption{Common parameters} \label{tab_parset}
\centering
\begin{tabular}{c|c|c}
\Xhline{1.2pt}
\Gape[2pt]{bandwidth} $W$ & number of subcarriers $N$ & power $P$  \\
\hline
\Gape[2pt]{5 MHz}& 512 &10 W \\
\Xhline{1.2pt}
\Gape[2pt]{carrier frequency} $f_c$ & train velocity $v$& path loss exponent $\alpha$  \\
\hline
\Gape[1pt]{3 GHz} & 100 m/s & 3 \\
\Xhline{1.2pt}
\Gape[2pt] {noise density $N_0$} & cell radius $R$ & vertical distance $d_v$ \\
\hline
 $6.32\times 10^{-16}$ W/Hz  & 5 km & 1 km\\
\Xhline{1.2pt}
 number of users $M$ & scheduling period & stepsize $\beta_{\text{sp}}$ \\
\hline
50 & 1 s & $10^{-3}$ \\
\Xhline{1.2pt}
\end{tabular}
\end{table}

\begin{table}[!ht]
\setlength{\abovecaptionskip}{2pt}
\setlength{\belowcaptionskip}{0pt}
\centering
\caption{Tap delays} \label{tab_tdelay}
\begin{tabular}{c|c|c|c|c|c|c}
\Xhline{1.2pt}
tap index & 1 & 2 & 3 & 4 & 5 & 6 \\
\hline
delay ($\mu$s) & 0 & 1 & 2 & 3 & 4 & 5\\
\hline
power & 1.000 &   0.368 & 0.135 & 0.050 & 0.018 & 0.007 \\
\Xhline{1.2pt}
\end{tabular}
\end{table}



\begin{figure}
  \centering
  \includegraphics[width=0.45\textwidth]{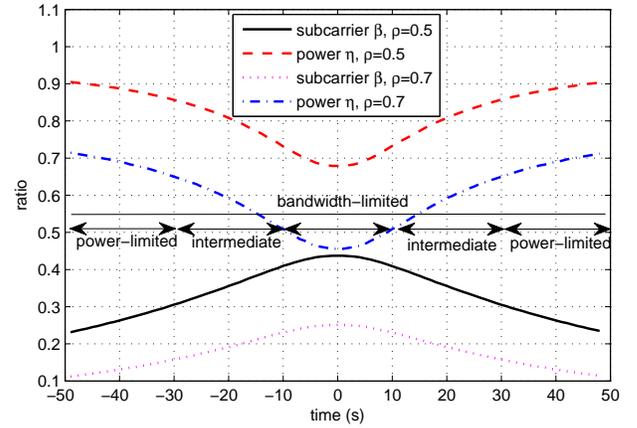}\\
  \caption{OPSA for the MR with ICI.}\label{fig_pwrsubcarICI}
\end{figure}

\begin{figure}
  \centering
  \includegraphics[width=0.45\textwidth]{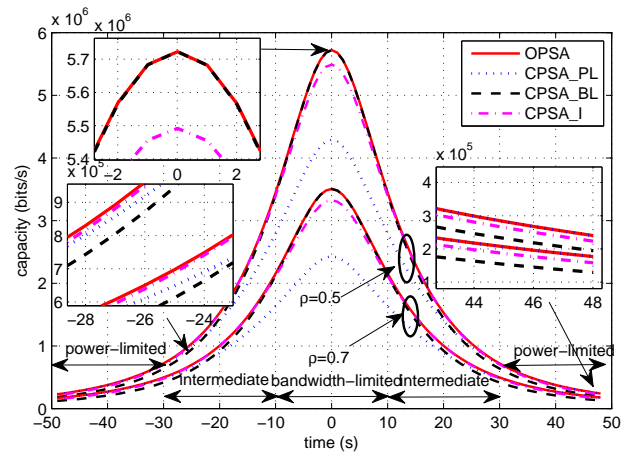}\\
  \caption{Capacity of the MR with ICI.}\label{fig_capICI}
\end{figure}


\subsubsection{Power and Subcarrier Allocation}
Fig. \ref{fig_pwrsubcarICI} shows power and subcarrier allocation of OPSA. One can see that as the MR comes closer to the BS from the cell edge, the power allocated to it decreases, while the number of subcarriers allocated to it increases. When the MR is far from the BS, the received SNR is small and the MR is power-limited. Therefore, more power but less subcarriers are needed to maximize its capacity. As the MR approaches the BS, the received SNR increases. Correspondingly, the efficiency of power decreases while that of bandwidth increases. Thus, it would be better for the MR to trade some of its power for subcarriers from local users. This process reverses when the MR travels far from the BS. Moreover, one can see that when the sum capacity constraint of local users decreases, both the power and subcarriers allocated to the MR increase.

\subsubsection{Achievable Capacities}
The capacity versus time curve of the MR obtained using Alg. \ref{alg_optsol} is shown in Fig. \ref{fig_capICI}. For comparison, we also plot three constant power and subcarrier allocation policies. CPSA-PL, CPSA-BL, CPSA-I are the policies with $\{\eta(i), \beta(i)\}$ chosen at the time when the train is farthest from the BS, nearest to the BS and somewhere which lies between. From this figure, one can see that CSPA-PL is approximately optimal when the MR is in the power-limited region. However, when the MR leaves that region, it works quite poor compared with the OPSA. CSPA-BL is approximately optimal in the bandwidth-limited region, and performs bad in the power-limited region. CSPA-I achieves the near optimal performance in the intermediate region, while in the other two regions, it has a performance which lies between CPSA-PL and CPSA-BL. Nevertheless, this figure explicitly shows the advantage of OPSA over CPSA.
%

\subsubsection{Normalized Capacity Gap}

\begin{figure}
  \centering
  \includegraphics[width=0.45\textwidth]{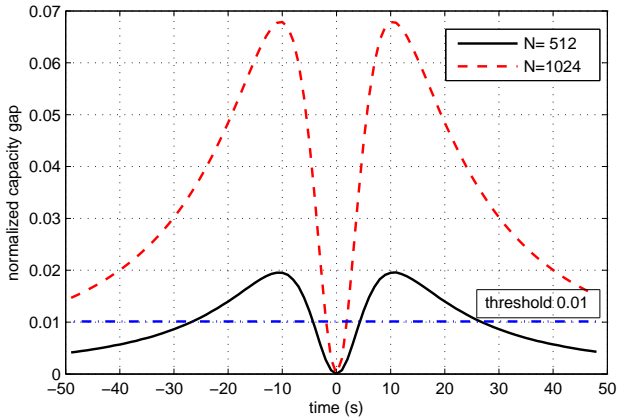}\\
  \caption{Normalized capacity gap for the MR using OPSA.}\label{fig_normrgap}
\end{figure}
The capacity obtained by treating ICI as Gaussian noise is the lower bound of the achievable rate for the MR. We can also obtain the upper bound by ignoring the ICI term in   (\ref{eqn_simmainopt}). Then practical ICI cancellation would obtain rates between these two bounds. By analyzing the gap between these two bounds, we can find the possible rate gain of ICI cancellation. We define the normalized capacity gap of those two bounds as  $C_{\text{gap}}=\frac{C_{\text{upper}}-C_{\text{lower}}}{C_{\text{upper}}}$ and  plot this gap in Fig. \ref{fig_normrgap}. We see that the normalized capacity gap is time-varying. It indicates that ICI cancellation, like schemes in \cite{zhang2004annicofdms} \cite{zhang2004icofdmsonbts}, is not always meaningful since the rate gain is little in regions where the normalized capacity gap is small, especially when the complexity is considered.

\section{Conclusion} \label{sec_conclusion}

In this paper, we investigated downlink resource allocation for a high-speed train and local users in an OFDMA system. Considering the differences between the channels of the MR in the train and local users, we allocated resources for them over two time-scales. We derived a closed-form expression for the ICI term at the MR. Then we transformed the problem into a convex optimization problem and provided an efficient algorithm to find the unique optimal solution. The performance of OPSA is compared with that of CPSA via simulations. Moreover, by comparing the capacity bounds with and without ICI, we found that ICI cancellation is not always necessary considering its complexity, it is worthy only when the possible rate gain is large.

\section*{Acknowledgement}

This work was partly supported by the China Major State Basic Research Development Program (973 Program) No.2012CB316100(2), National Natural Science Foundation of China (NSFC) No.61171064 and NSFC No. 61021001, where Ke xiong's work is  supported by NSFC No. 61201203.

\bibliographystyle{ieeetr}
\bibliography{pwrsubcaralloc}

\end{document}